%% file: main.tex
\documentclass[authoryear,preprint,12pt]{elsarticle}

\usepackage{amsmath}
\usepackage{amssymb}
\usepackage{graphicx}
\usepackage{booktabs}
\usepackage{multirow,array}
\usepackage{placeins}
\usepackage{url}
\usepackage[hidelinks]{hyperref}
\makeatletter
\let\ps@pprintTitle\ps@plain
\makeatother
\graphicspath{{figures/}}
\begin{document}

\begin{frontmatter}

\title{Reconstruction quality and restart repeatability are insufficient indicators
of muscle synergy recovery by nonnegative matrix factorisation}

\author[aff1]{Ye Ma}
\author[aff2]{Dongwei Liu\corref{cor1}}
\ead{dongwei.liu@zufe.edu.cn}
\author[aff3]{Meijin Hou\corref{cor2}}
\ead{meijin.hou@fjmu.edu.cn}
\author[aff4]{Chenyi Guo}
\cortext[cor1]{Corresponding author}
\cortext[cor2]{Corresponding author}

\affiliation[aff1]{
  organization={Research Academy of Grand Health, Faculty of Sports Sciences, Ningbo University},
  addressline={818 Fenghua Road},
  city={Ningbo},
  postcode={315211},
  country={China}
}
\affiliation[aff2]{
  organization={School of Information Technology and Artificial Intelligence, Zhejiang University of Finance and Economics},
  addressline={No. 18 Xueyuan Road},
  city={Hangzhou},
  postcode={310018},
  country={China}
}
\affiliation[aff3]{
  organization={National and Local Joint Engineering Research Centre of Rehabilitation Medical Technology; College of Rehabilitation Medicine, Fujian University of Traditional Chinese Medicine},
  addressline={No. 1 West Qiuyang Road},
  city={Fuzhou},
  postcode={350122},
  country={China}
}
\affiliation[aff4]{
  organization={Department of Electronic Engineering, Tsinghua University},
  addressline={No. 100 Xueyuan Road},
  city={Beijing},
  postcode={100084},
  country={China}
}

\begin{abstract}
Muscle-synergy studies use factorisation methods such as nonnegative matrix
factorisation (NMF) to extract muscle synergies from surface electromyography (sEMG).
Reconstruction quality and restart repeatability are commonly used to evaluate these
estimates, but whether they indicate recovery of underlying synergies remains
unclear. Here, we defined synergy recovery as selection of the correct ground-truth
rank and a minimum matched spatial cosine similarity of at least 0.80. We generated
4320 independent 16-muscle sEMG sequences across five simulation factors:
ground-truth rank, signal-to-noise ratio, sequence length, spatial mixing and
activation mixing. Rank selection used a variance accounted for (VAF) threshold of
0.90. At the selected ranks, NMF achieved VAF $\geq 0.90$ for every sequence,
with median restart repeatability of 0.985. However, synergy recovery was achieved
in only 7.92\% of sequences. Supplying the ground-truth
rank increased recovery to 60.79\%. Reconstruction quality and restart repeatability
had modest associations with spatial cosine similarity (Spearman's $\rho = 0.253$ and
$0.281$, respectively). Recovery varied with all five factors, with nine significant
two-factor interactions in exploratory analyses. In a supporting analysis, we
assessed recovery using reference signals constructed from spatial weights and
activation coefficients estimated from measured sEMG during five movement conditions.
Selected ranks were lower than reference ranks in all five conditions despite high
reconstruction quality. Accurate reconstruction and repeatable estimates therefore
did not establish recovery of the underlying synergy structure. Physiological
interpretations require additional evidence of functional relevance and assessment of
sensitivity to the selected rank.
\end{abstract}

\begin{keyword}
muscle synergies \sep nonnegative matrix factorisation \sep factorial simulation
\sep rank-selection accuracy
\sep synergy recovery \sep restart repeatability
\end{keyword}
\end{frontmatter}

\section{Introduction}
\label{sec:introduction}

Muscle synergy studies use surface electromyography (sEMG)
to analyse coordinated muscle activity \citep{clark2010}.
Estimated synergies have been used to characterise coordination complexity
\citep{clark2010}, neurological impairment \citep{cheung2012,banks2017},
motor learning \citep{park2022}, fatigue \citep{chen2024},
and post-stroke motor recovery
\citep{hashiguchi2016}.

Several factorisation methods have been used to extract muscle synergies
\citep{tresch2006,ebied2018}.
Nonnegative matrix factorisation (NMF) is commonly used 
because it represents sEMG envelopes as additive combinations
of nonnegative components,
which is consistent with the physiological interpretation of sEMG envelope
amplitudes as nonnegative indicators of muscle activation \citep{lee1999,tresch2006}.

NMF approximates a multichannel sEMG envelope
matrix $V$ by the product $\widehat W_r\cdot\widehat C_r$
at a candidate rank $r$ (the number of fitted components).
The component number is commonly selected using a {\it variance accounted for} (VAF) threshold,
the elbow of the VAF curve, or other model-selection criteria
\citep{tresch2006,ranaldi2021}.
Each column of the spatial weight matrix
$\widehat W_r$ is a muscle-weight vector, while the corresponding row of the activation
coefficient matrix $\widehat C_r$ describes its time-varying activation coefficients \citep{tresch2006,muceli2010}.

Two commonly used measures for assessing NMF decompositions are reconstruction
quality and restart repeatability.
\emph{Reconstruction quality}, commonly quantified by VAF,
describes how closely $\widehat W_r\cdot\widehat C_r$ approximates the input envelopes $V$
\citep{tresch2006,clark2010}.
\emph{Restart repeatability} describes the consistency of $\widehat W_r$
across optimisations from different initialisations for a fixed input and rank
\citep{shourijeh2016}.

Muscle synergy analysis aims to investigate how the central nervous system
(CNS) organises muscle coordination.
High reconstruction quality indicates that the estimated model closely reproduces
the input sEMG envelopes, whereas high restart repeatability indicates consistent
spatial estimates across initialisations. 
Neither measure alone establishes that
the estimated synergies represent the underlying neural organisation of muscle
coordination \citep{tresch2006,delis2013,kutch2012}.


Ground-truth synergies are unknown in measured sEMG. Simulations with known
spatial weights and activation coefficients therefore provide a controlled
setting to assess how accurately NMF identifies the underlying synergy
structure. We synthesised sEMG envelope matrices from a spatial weight matrix
$W$, an activation coefficient matrix $C$ and a disturbance matrix $n$.
We then compared the selected component number and estimated muscle-weight
vectors with their ground-truth counterparts. Spatial agreement was quantified
by spatial cosine similarity between matched estimated and ground-truth
muscle-weight vectors. In this study, we defined synergy recovery as meeting
two conditions: the evaluated rank equalled the ground-truth rank, and the
spatial cosine similarity of every matched vector pair was at least 0.80.

Synergy estimates can be influenced by noise \citep{delis2013}, sequence length
\citep{oliveira2014}, the number of underlying muscle synergies
\citep{burkholder2013}, and spatial similarity among muscle-weight vectors or
correlations among activation coefficients \citep{soomro2018}. We systematically
varied five corresponding simulation factors to generate 4320 independent
16-muscle sEMG sequences in a factorial design.

The study had three objectives: (1)~to evaluate synergy recovery of
NMF with a specified VAF-based rank-selection rule; (2)~to assess the extent
to which reconstruction quality and restart repeatability indicate spatial agreement 
between estimated and generating muscle-weight vectors, 
quantified by spatial cosine similarity; 
and (3)~to assess synergy recovery across diverse signal conditions 
by systematically varying five factors 
and examining their two-factor interactions. 
We also examined whether the simulation findings held for signals constructed
from measured sEMG across five movement conditions,
level, ramp and stair walking \citep{camargo2021}, 
and Tai Chi Heel Kick with left and right stance.

\section{Methods}
\begin{figure}[!tbp]\centering
\includegraphics[width=\textwidth,height=.76\textheight,keepaspectratio]{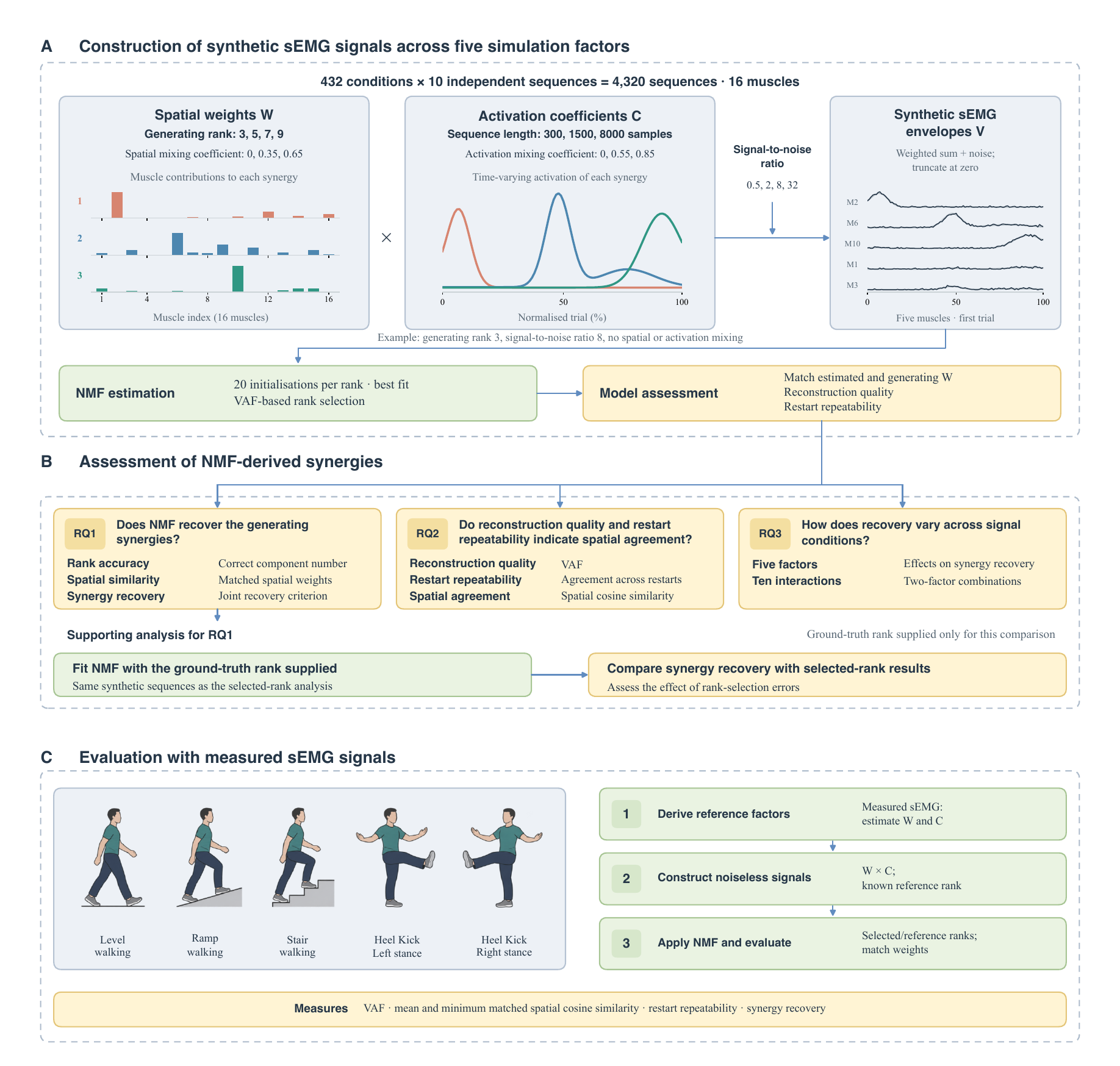}
{\renewcommand{\baselinestretch}{1}\small
\caption{Study design. (A) Synthetic surface electromyography (sEMG) 
construction across five factors: ground-truth rank, spatial mixing, sequence length, activation mixing 
and signal-to-noise ratio. The design comprised 432 conditions with ten independent sequences each. 
Colours identify corresponding components in spatial weights $W$ and activation coefficients $C$. 
(B) Three research questions (RQ1--RQ3) assess synergy recovery, 
associations of reconstruction quality and restart repeatability with spatial agreement, 
and recovery across signal conditions. 
A supporting analysis for RQ1 compares recovery at selected and supplied ground-truth ranks. 
(C) Evaluation of noiseless reference signals constructed from factors estimated 
from measured sEMG across five movement conditions. 
NMF: nonnegative matrix factorisation; VAF: variance accounted for.}

\label{fig:design}}\end{figure}

We used a factorial simulation to assess rank-selection accuracy and synergy
recovery, examine the associations of reconstruction quality and
restart repeatability with spatial similarity, and determine the effects of five
simulation factors. Supporting analyses evaluated the same sequences with the
ground-truth rank supplied and examined signals constructed from sEMG-derived
reference factors (Fig.~\ref{fig:design}).

\subsection{Simulation design and signal construction}
Each sequence contained $M=16$ muscles and $L\in\{3,15,80\}$ trials of
100 time-normalised samples, giving $T=100L$ samples. The sEMG envelope
matrix $V\in\mathbb R_+^{M\times T}$ was the input to NMF. 
Each column of $W\in\mathbb R_+^{M\times k}$ represented a ground-truth spatial weight
vector, and the corresponding row of $C\in\mathbb R_+^{k\times T}$ contained
its time-varying activation coefficients. The ground-truth rank $k$ was the
number of synergies. We constructed the signals as
\begin{equation}
V=WC+n,
\label{eq:signal-generation}
\end{equation}
where $n$ includes the added Gaussian noise and the effect of setting negative
values to zero. Spatial weights were shared across trials, while activation
coefficients varied in amplitude, timing and duration.

The full factorial design crossed ground-truth rank $k\in\{3,5,7,9\}$,
signal-to-noise ratio $\eta\in\{0.5,2,8,32\}$, sequence length
$T\in\{300,1500,8000\}$, spatial mixing coefficient $b_W\in\{0,0.35,0.65\}$
and activation mixing coefficient $b_C\in\{0,0.55,0.85\}$. Spatial mixing
increased similarity among muscle-weight vectors, while activation mixing
increased overlap among activation coefficients. Signal-to-noise ratio was
the linear ratio of signal variance to noise variance before truncation at
zero. Ten independently generated sequences per condition yielded
4320 sequences across 432 conditions. Sampling distributions, normalisation,
mixing procedures and noise construction are specified in
\ref{app:generator}.

\subsection{NMF estimation and rank selection}
For each sequence, NMF estimated nonnegative spatial weights $\widehat W_r$
and activation coefficients $\widehat C_r$ by minimising reconstruction sum
of squared errors (SSE). We evaluated candidate ranks $r=1,\ldots,16$, with
20 initialisations per rank, and retained the lowest-SSE fit at each rank.
Ground-truth information was not used to select a fit or rank. Optimisation
used coordinate descent in scikit-learn \citep{pedregosa2011}; initialisation
and stopping settings are specified in \ref{app:estimation}.

Reconstruction quality was quantified separately for each sequence using
globally centred variance accounted for (VAF). The mean was calculated across
all muscles and time points within that sequence, including its concatenated
trials. VAF was calculated from the lowest-SSE fit at each candidate rank
against the noisy input:
\begin{equation}
v_r=1-\frac{\|V-\widehat W_r\widehat C_r\|_F^2}
{\|V-\bar V\mathbf1\|_F^2},
\label{eq:vaf-wc}
\end{equation}
where $\bar V$ is this sequence-specific mean, $\mathbf1$ is an all-ones matrix
of the same dimensions as $V$, and $\|\cdot\|_F$ is the Frobenius norm.
The selected rank $\widehat k$ was the smallest rank achieving VAF
$\geq0.90$, or 16 if none met the threshold. Reconstruction quality and
spatial assessment used the same lowest-SSE fit.

\subsection{Assessment of synergy recovery}
To address RQ1, we compared the selected rank and estimated muscle-weight
vectors with the ground truth. Rank-selection accuracy was the percentage of
sequences with $\widehat k=k$; lower and higher selected ranks indicated
under-selection and over-selection, respectively.

Estimated and ground-truth spatial weight vectors were paired by optimal one-to-one assignment to
maximise total spatial cosine similarity \citep{kuhn1955}. We summarised
spatial agreement using the mean and minimum similarity across matched pairs.
When the selected and ground-truth ranks differed, only $\min(k,\widehat{k})$
pairs were matched, and unmatched
components were excluded from these summaries. Synergy recovery required
both the correct ground-truth rank and a minimum matched spatial cosine
similarity of at least 0.80, informed by published spatial-similarity criteria
\citep{saito2018,kaufmann2024synergies}. Recovery percentages included all
4320 sequences and assessed component number and spatial weights, not
activation coefficients. Definitions and matching details are given in
\ref{app:assessment}.

As a supporting analysis, we repeated the assessment with the ground-truth
rank supplied, using the same fitting and matching procedures, to examine
recovery without rank-selection errors.

\subsection{Reconstruction quality, restart repeatability and spatial agreement}
To address RQ2, we examined whether VAF and restart repeatability indicated
spatial agreement with the ground truth. Restart repeatability measured
agreement among the spatial estimates from the 20 initialisations at the
selected rank. Normalised muscle-weight vectors were pooled and clustered
to form a consensus; each initialisation was matched to this consensus, and
the matched spatial cosine similarities were averaged across vectors and
initialisations. The consensus was used only for restart repeatability,
not for reconstruction or ground-truth comparisons
(\ref{app:estimation} and~\ref{app:assessment}).

Spearman's $\rho$ quantified associations of VAF and restart repeatability
with mean spatial cosine similarity. 
Linear regressions described the pooled trends. 
Additional regressions after subtracting condition means examined
associations within conditions (\ref{app:association-details}).

\subsection{Factor effects and statistical analysis}
Each independent sequence was one observation. We reported rank-selection
outcomes and recovery as counts and percentages, and continuous measures as
medians and 25th--75th percentiles. Repeatability summaries and associations
used valid estimates only. Confidence intervals were obtained by resampling
sequences within conditions, preserving the factorial design.

To address RQ3, we assessed the five main effects and ten two-factor
interactions on synergy recovery. Recovery proportions were calculated from
the ten sequences in each condition; factor-level percentages averaged
equally across the remaining factors. Wald tests used condition proportions
and their estimated variances. Benjamini--Hochberg adjustment covered the
15 exploratory tests; pairwise comparisons following significant main effects
used Holm adjustment within each factor. Adjusted $p$ values below 0.05 were
considered significant. Bootstrap settings, test calculations and uncertainty
estimation are specified in \ref{app:association-details}
and~\ref{app:factor-statistics}.

\subsection{Supporting analysis using measured sEMG}
We assessed whether the findings extended to reference signals constructed
from measured sEMG during five movement conditions: level, ramp and stair
walking, and Tai Chi Heel Kick with left and right stance. Walking signals
contained ten muscles and six, six and seven reference synergies,
respectively; Tai Chi signals contained 16 muscles and six reference synergies
for each stance.

Spatial weights were estimated from measured sEMG and held fixed while
nonnegative least squares estimated activation coefficients. Their product
formed noiseless reference signals with known reference ranks. We applied
the same NMF estimation and rank-selection procedures over ranks 1--10 for
walking and 1--16 for Tai Chi, and assessed reconstruction quality, spatial
agreement, restart repeatability and synergy recovery. A supplied-reference-rank
analysis assessed recovery without rank-selection errors. Reference-signal
construction and assessment procedures are detailed in
\ref{app:reference-signals}.

\section{Results}
\subsection{Reconstruction quality and restart repeatability}

All 4320 sequences achieved VAF $\geq0.90$ at the selected rank.
Median VAF was 0.915
(25th--75th percentiles: 0.907--0.925). 
Restart repeatability was calculable
for all sequences, with a median of 0.985
(25th--75th percentiles: 0.962--0.999).

\subsection{Rank-selection accuracy and synergy recovery}

\begin{figure}[!tbp]\centering
\includegraphics[width=\textwidth,height=0.62\textheight,keepaspectratio]{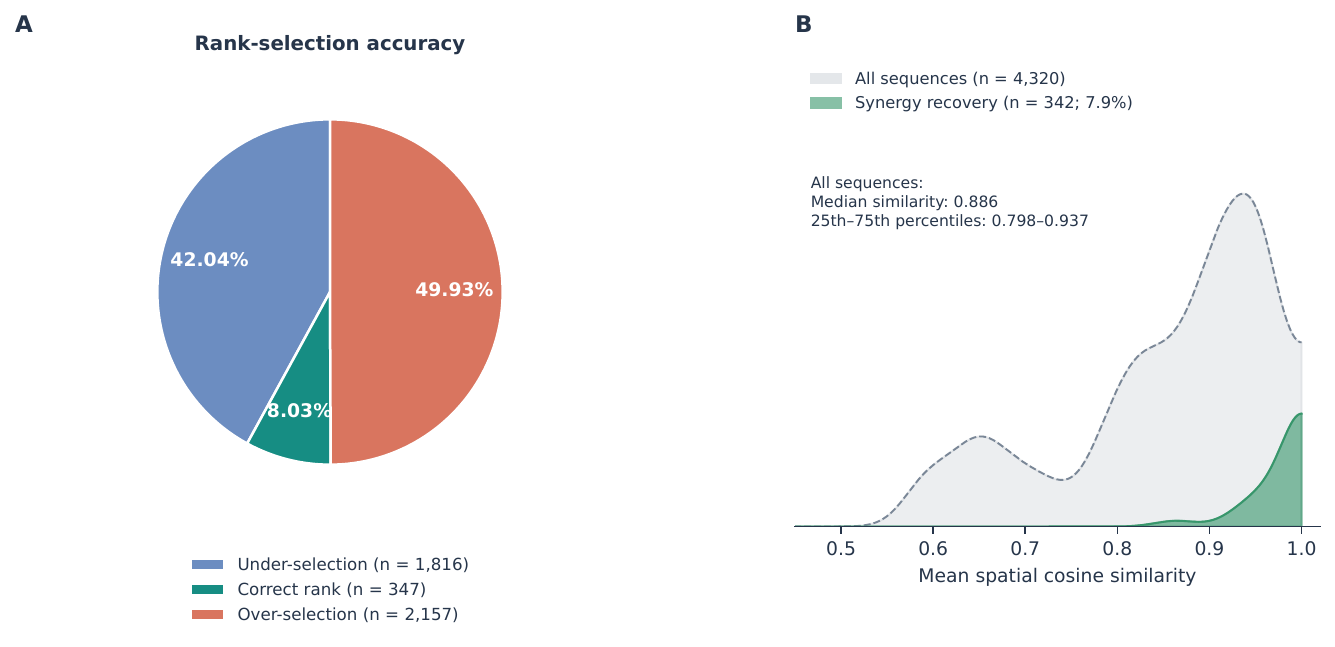}
{\renewcommand{\baselinestretch}{1}\small
\caption{Rank-selection accuracy, spatial similarity and synergy recovery.
(A) Percentages of sequences with under-selected, correct and over-selected
ranks. (B) Distribution of mean spatial cosine similarity across all 4320
sequences, with the contribution of sequences achieving synergy recovery
shaded green.}
\label{fig:rq1-selector}}
\end{figure}

Across the 4320 synthetic sequences, 
the selected rank matched the ground-truth rank 
exactly in 347 sequences (8.03\%), 
was lower than the ground-truth rank in 1816 (42.04\%) and was higher in 2157
(49.93\%; Fig.~\ref{fig:rq1-selector}A).
The median of the sequence-level mean spatial
cosine similarities between matched estimated and ground-truth muscle-weight
vectors was 0.886 (25th--75th percentiles: 0.798--0.937; 
Fig.~\ref{fig:rq1-selector}B).
Synergy recovery,
requiring exact selection of the ground-truth rank 
and a minimum matched spatial cosine similarity of at least 0.80 
between estimated and ground-truth muscle-weight vectors,
was achieved in 342/4320 sequences (7.92\%; 95\% CI: 7.64--8.19\%).

Supplying the ground-truth rank increased synergy recovery from
342/4320 (7.92\%) to 2626/4320 (60.79\%). 
Even with the correct rank, 1694 sequences (39.21\%) had a minimum matched
spatial cosine similarity below 0.80.

Across the five sEMG-derived reference conditions, 
the selected rank was five in every condition, 
below the reference rank of six or seven, 
while VAF exceeded 0.90 in all conditions.
Table~\ref{tab:real-semgrecovery} summarises the selected number of synergies,
reconstruction quality, spatial cosine similarity and restart repeatability.

\begin{table}[!tbp]\centering
\renewcommand{\baselinestretch}{1}\small
\caption{Rank selection and reconstruction of signals constructed from sEMG-derived reference factors.}
\label{tab:real-semgrecovery}
\vspace{5pt}
\setlength{\tabcolsep}{5pt}\renewcommand{\arraystretch}{1.12}
\begin{tabular}{@{}lrrrrrr@{}}\toprule
Condition & $k_{\mathrm{ref}}$ & $\widehat k$ & VAF & $s_W$ & $s_W^{\min}$ & $A_W$\\\midrule
Level walking & 6 & 5 & 0.9514 & 0.9781 & 0.9078 & 0.9999\\
Ramp walking & 6 & 5 & 0.9327 & 0.9694 & 0.8903 & 0.9408\\
Stair walking & 7 & 5 & 0.9182 & 0.9768 & 0.9427 & 1.0000\\
Tai Chi DJ-L & 6 & 5 & 0.9329 & 0.9906 & 0.9781 & 0.9552\\
Tai Chi DJ-R & 6 & 5 & 0.9397 & 0.9812 & 0.9300 & 1.0000\\
\bottomrule\end{tabular}
\par\vspace{5pt}\begin{minipage}{\textwidth}\footnotesize
\textit{Notes.} $k_{\mathrm{ref}}$: number of reference synergies used to
construct each signal; $\widehat k$: selected number of synergies;
VAF: variance accounted for; $s_W$: mean spatial cosine similarity;
$s_W^{\min}$: minimum matched spatial cosine similarity. Both spatial summaries
compare matched estimated and reference muscle-weight vectors;
$A_W$: restart repeatability across 20 NMF initialisations.
DJ-L/R: Heel Kick with left/right stance.
\end{minipage}\end{table}

\subsection{Associations of reconstruction quality and restart repeatability with spatial agreement}

\begin{figure}[!tbp]\centering
\includegraphics[width=\textwidth,height=0.6\textheight,keepaspectratio]{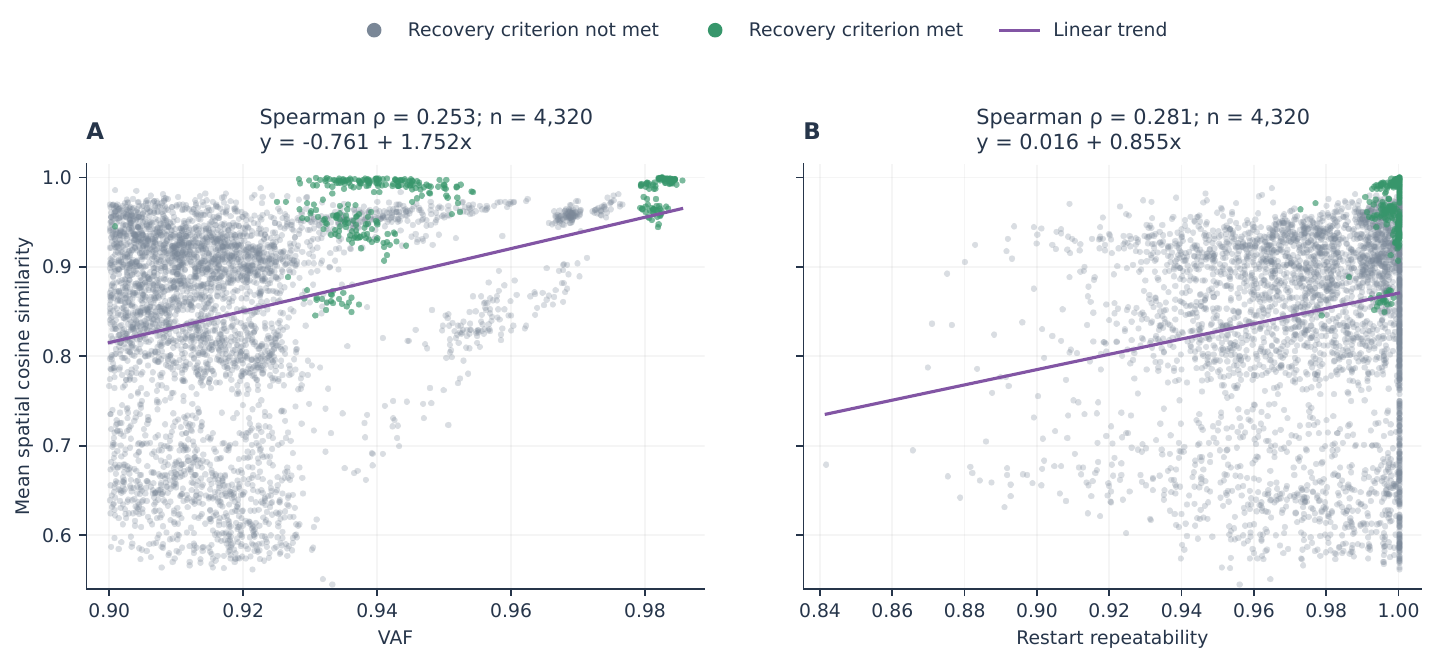}
{\renewcommand{\baselinestretch}{1}\small
\caption{Associations of reconstruction quality and restart repeatability
with spatial similarity. Mean spatial cosine similarity is plotted against
(A) VAF and (B) restart repeatability. Each point represents one
simulated sequence; green points indicate synergy recovery. Purple lines show
linear fits, and Spearman's $\rho$ values summarise the associations.}
\label{fig:metrics}}\end{figure}

VAF and restart repeatability had modest positive associations with
mean spatial cosine similarity across all 4320 sequences (Spearman's
$\rho=0.253$ and $0.281$, respectively; Fig.~\ref{fig:metrics}A--B).
Within simulation conditions, the VAF association remained positive, whereas
restart repeatability showed little association with spatial similarity.

\subsection{Factor effects on synergy recovery}
\begin{figure}[!tbp]\centering
\includegraphics[width=\textwidth,height=.66\textheight,keepaspectratio]{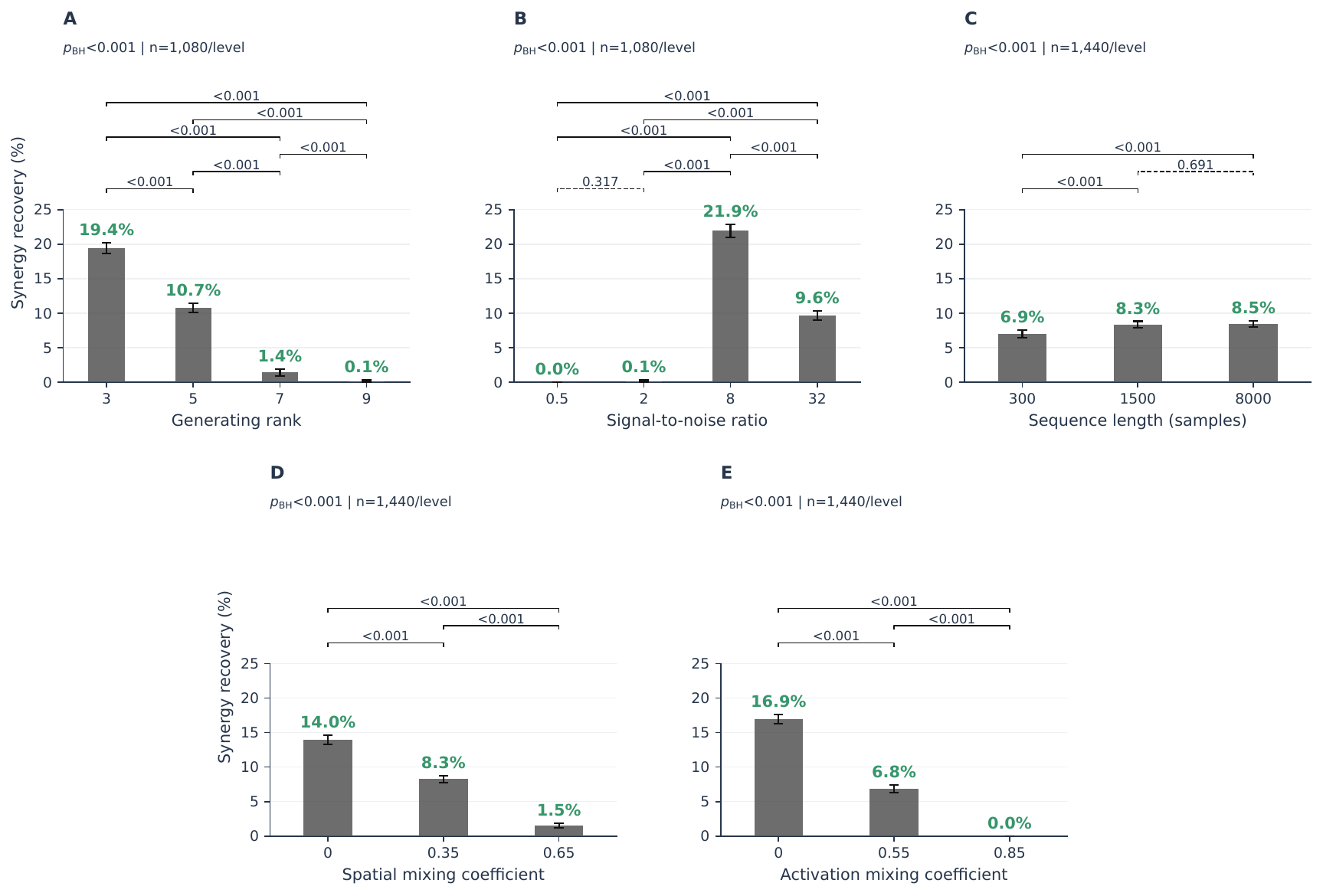}
{\renewcommand{\baselinestretch}{1}\small
\caption{Effects of five simulation factors on synergy recovery.
Bars show recovery percentages averaged across the other four factors;
error bars indicate 95\% confidence intervals. Panel annotations report
BH-adjusted $p$-values for factor effects. Brackets show Holm-adjusted pairwise
comparisons, with dashed brackets indicating $p\geq0.05$.}
\label{fig:rq3-recovery}}\end{figure}

Synergy recovery differed across the levels of all five factors
in the exploratory BH-adjusted Wald tests 
(all $p_{\mathrm{BH}}<0.001$). 
Synergy recovery was summarised by factor level,
pooling sequences across all levels
of the remaining four factors (Fig.~\ref{fig:rq3-recovery}).

Synergy recovery decreased as ground-truth rank increased: 19.4\%, 10.7\%, 1.4\% and
0.1\% at ranks 3, 5, 7 and 9, respectively. 
Increasing either mixing coefficient also reduced synergy recovery.
Synergy recovery declined from 14.0\% to 8.3\% and 1.5\% at
spatial mixing coefficients of 0, 0.35 and 0.65, 
and from 16.9\% to 6.8\% and 0\% at activation mixing coefficients 
of 0, 0.55 and 0.85. 
All pairwise comparisons within these three factors were significant
($p_{\mathrm{Holm}}<0.001$).

Synergy recovery varied non-monotonically with signal-to-noise ratio,
reaching 0\%, 0.1\%, 21.9\% and 9.6\% at ratios of 0.5, 2, 8 and 32, respectively. 
The first two levels did not differ significantly ($p_{\mathrm{Holm}}=0.317$); 
all remaining pairwise comparisons were significant ($p_{\mathrm{Holm}}<0.001$).
Increasing sequence length produced a smaller change: 
synergy recovery was 6.9\%, 8.3\% and 8.5\% at 300, 1500 and 8000 samples,
with no significant difference between 1500 and 8000 samples ($p_{\mathrm{Holm}}=0.691$).

Nine of the ten two-factor interactions were significant
($p_{\mathrm{BH}}<0.001$); 
the exception was sequence length $\times$ spatial mixing coefficient ($p_{\mathrm{BH}}=0.0503$). 
For the spatial $\times$ activation mixing interaction, 
synergy recovery decreased from 26.7\% to 4.6\% as
spatial mixing increased from 0 to 0.65 while activation mixing remained at 0.
At the highest activation mixing coefficient, 0.85, synergy recovery was zero at every
spatial mixing level. These percentages average across the remaining three
factors. 
Detailed statistical results are provided in the appendix
(\ref{app:factor-statistics}).

\section{Discussion}
High reconstruction quality and restart repeatability did not ensure recovery
of the ground-truth muscle synergies. In this simulation, the ground-truth
factors represented the synergies assumed to underlie central nervous system
(CNS) control of muscle coordination. Although all 4320 selected models achieved
VAF $\geq 0.90$ and median restart repeatability was 0.985, only 342 sequences
(7.92\%) met the synergy recovery criterion. The central finding is therefore
a distinction between reconstructing sEMG signals and recovering their
underlying synergy structure: an NMF model can reconstruct signals accurately
and produce consistent estimates across initialisations without recovering the
components specified in the ground-truth model.

\subsection{Reconstruction quality does not establish synergy recovery}
Signal reconstruction and synergy recovery evaluate different properties of an
estimated model. Reconstruction quality measures agreement between the observed
and reconstructed sEMG envelopes. More generally, recovering the underlying
synergies requires identifying the number and spatial organisation of the
components responsible for the observed muscle activity. A high VAF indicates
that the estimated components, when combined, reproduce much of the variation
in the recorded signals. It does not establish that the number of components
or the muscle weights within each component correspond to the underlying
synergies. Our simulation demonstrates this distinction by comparing the
estimated components with a known ground-truth structure.

Most recovery failures involved a mismatch between the selected and ground-truth
ranks. The selected rank was correct in only 347 sequences (8.03\%), of which
342 also met the spatial similarity criterion. Under-selection and over-selection
occurred in 42.04\% and 49.93\% of sequences, respectively. The VAF-based rule
selects the smallest component number that meets the reconstruction threshold.
In measured sEMG, the ground-truth component number is unknown, so this threshold
provides a reconstruction-based selection criterion rather than direct
verification of the underlying synergy number. In the supporting analysis,
supplying the ground-truth rank increased recovery to 60.79\%, indicating that
rank mismatch was a major constraint, although the correct rank alone was
insufficient to ensure recovery. Previous simulations showed that
component-number selection depends on noise characteristics, and that spatial
estimation depends on noise and correlations between activation coefficients
\citep{tresch2006}. Spatial estimation also depends on synergy sparsity and the
number of recorded channels \citep{ebied2018}. Comparisons with alternative selection methods
further indicate that the selected rank depends on the criterion used
\citep{delis2013,ranaldi2021}.

To examine whether this finding extended to signal structures derived from
measured sEMG, the supporting analysis evaluated five movement conditions:
level, ramp and stair walking, and Tai Chi Heel Kick with left and right stance.
For each condition, a noiseless reference signal was constructed from factors
estimated from measured sEMG. All five conditions were under-selected despite
high reconstruction quality and minimum matched spatial cosine similarities
above 0.89. Each selected model missed one or two reference components, whereas
supplying the reference rank allowed all five models to meet the recovery
criterion. Thus, the estimated components could reconstruct the reference
signals accurately and closely match individual reference muscle-weight vectors
while still failing to recover the full set of reference synergies.

\subsection{Restart repeatability does not establish agreement with ground-truth synergies}
Restart repeatability describes consistency across repeated initialisations,
rather than correspondence with the ground-truth factors. Its high median value
indicates that repeated initialisations generally produced consistent spatial
estimates. However, agreement among estimates does
not establish agreement with the ground-truth synergies. Restart repeatability
measures consistency when fitting the same input at the same rank. It is distinct
from consistency across gait cycles \citep{kim2016} or similarity between
participants \citep{shourijeh2016}.

Reconstruction quality and restart repeatability had positive but modest
associations with mean spatial cosine similarity, with Spearman correlations of
0.253 and 0.281, respectively. Neither measure closely tracked spatial similarity
across the simulated sequences. Within simulation conditions, restart
repeatability showed little association with spatial similarity, indicating
that the positive pooled association did not describe a clear relationship
within conditions.

Mean spatial cosine similarity nevertheless showed substantial agreement between
matched estimated and ground-truth muscle-weight vectors: its median across
sequences was 0.886, with 25th--75th percentiles of 0.798--0.937. This result is
compatible with the low recovery percentage because the measures answer
different questions. Mean spatial cosine similarity summarises agreement across
matched components, whereas the recovery criterion used here additionally
requires the correct ground-truth rank and adequate spatial cosine similarity
for every matched component. An average similarity can therefore conceal an
incomplete component set or poorer recovery of individual components.

These distinctions matter when interpreting NMF-derived components as evidence
of CNS organisation. Within the simulation, the assumed ground-truth synergies
were known, making their recovery directly testable. In measured sEMG,
physiological interpretation requires evidence linking the estimated components
to the proposed coordination mechanisms \citep{ting2015,kutch2012,delis2013}.
Such evidence should address whether the components explain coordinated muscle
activity in relation to task demands or biomechanical function, and whether
that interpretation remains consistent across relevant experimental conditions.
This links the proposed physiological role to observable behaviour beyond the
quality of signal reconstruction.

Compact sEMG representations have practical applications in myoelectric
prosthesis control \citep{farina2014,jiang2009}, rehabilitation assessment
\citep{cheung2012,ting2015} and postural-control analysis \citep{ting2007}.
Evaluation against functional outcomes can establish their value for a specific
application, even when recovery of the ground-truth synergies is not demonstrated.
Reporting reconstruction quality, restart repeatability and sensitivity to the
selected rank alongside evidence of physiological relevance would help readers
assess those interpretations.

\subsection{Synergy recovery across diverse signal conditions}
Synergy recovery varied across diverse signal conditions and depended on their
combinations. Marginal recovery declined from 19.4\% at ground-truth rank 3 to
0.1\% at ground-truth rank 9. With the number of muscles fixed at 16, increasing
the ground-truth rank increased the number of muscle-weight vectors that had
to be selected and estimated simultaneously. This pattern is consistent with
evidence that spatial estimation depends on the number of recorded channels
relative to the component number \citep{ebied2018}. Because recovery required
both the correct rank and adequate spatial cosine similarity, these percentages
reflect the combined effects of rank selection and spatial estimation.

Increasing either mixing coefficient also reduced recovery. Recovery declined
from 14.0\% to 1.5\% across spatial mixing levels and from 16.9\% to zero across
activation mixing levels. Spatial mixing increased similarity among ground-truth
muscle-weight vectors, whereas activation mixing increased overlap among the
ground-truth activation coefficients. Both reduced differences between components
that may assist their separation. These findings agree with previous simulations
showing that correlated activations and task constraints can obscure ground-truth
synergies \citep{burkholder2013,steele2015}.

Increasing sequence length improved synergy recovery, with limited gains beyond 1500 samples. Recovery increased from 6.9\% at
300 samples to 8.3\% at 1500 samples and 8.5\% at 8000 samples. The latter two
levels did not differ significantly in the Holm-adjusted comparison
($p_{\mathrm{Holm}}=0.691$). Additional observations under the same simulation
model therefore offered limited improvement beyond 1500 samples. Previous work
has shown that gait-cycle count and averaging versus concatenation affect the
reconstruction of walking records \citep{oliveira2014}. Here, increasing sequence
length alone did not resolve the rank-selection and spatial-estimation errors,
suggesting that the information available to distinguish components mattered
alongside the number of observations.

The effect of signal-to-noise ratio was non-monotonic: recovery was absent at
0.5, peaked at 8 and declined at 32. At the lowest ratio, selected ranks of
12--14 exceeded every ground-truth rank in the design, preventing recovery under
the exact-rank criterion. The relationship between noise and recovery must
therefore be interpreted together with the behaviour of the VAF-based
rank-selection rule. It does not describe the effect of noise on spatial
estimation alone or imply that adding noise generally improves recovery.

Nine of the ten two-factor interactions were significant in the exploratory
BH-adjusted Wald tests. These results indicate that the effect of one factor
depended on the levels of other factors. For example, recovery was absent at the
highest activation mixing level across spatial mixing levels, illustrating how
strong activation overlap can constrain recovery despite changes in spatial
mixing. Marginal percentages therefore summarise outcomes averaged across the
other factors rather than performance in every combination of conditions.
Ground-truth rank, spatial and activation mixing, sequence length and noise
should consequently be considered jointly when interpreting synergy recovery.

\subsection{Limitations}
This study evaluated one NMF procedure, one VAF-based rank-selection rule and a
16-muscle simulation design. The recovery percentages and relationships across
signal conditions may differ with alternative estimation procedures,
rank-selection criteria or recording configurations.

\section{Conclusions}
Accurate reconstruction of sEMG signals and consistent NMF estimates across
initialisations did not ensure recovery of the underlying muscle synergies.
In simulations representing assumed CNS-generated synergies, only 7.92\% of
sequences met the recovery criterion despite uniformly high reconstruction
quality and high restart repeatability. Supplying the ground-truth rank
increased recovery to 60.79\%, showing that rank selection was a major constraint.
Reconstruction quality and restart repeatability had only modest associations
with spatial cosine similarity, while recovery varied across diverse signal
conditions and their interactions. Supporting analyses across five movement
conditions showed that signals constructed from measured-sEMG-derived factors
could also be reconstructed accurately with an incomplete set of reference
synergies. These findings indicate that reconstruction quality and restart
repeatability alone are insufficient evidence that NMF-derived components
reflect the underlying organisation of muscle coordination. Physiological
interpretations should be supported by evidence of functional relevance and
assessment of sensitivity to the selected rank.

\section*{Ethics statement}
This study used synthetic signals and existing sEMG data; no new participants
were recruited. The original Tai Chi recordings were collected under approval
from the Ethics Review Committee of the Rehabilitation Hospital Affiliated
to Fujian University of Traditional Chinese Medicine (2024KY-038-01).
All participants in that study provided written informed consent, and data
collection followed the Declaration of Helsinki and institutional guidelines.
Walking data came from the published dataset of \citet{camargo2021}.
\section*{Funding}
This work was supported by the National Natural Science Foundation of China
(grant nos. 12572368, 82405535 and 82575192);
the Natural Science Foundation of Fujian Province (grant no. 2026J001913);
the Fuzhou Science and Technology Project (grant nos. 2025-P-005 and 2025E3005);
the 2025 Ningbo University High-Level Science and Technology Project Incubation Program
(grant no. GJPY2025027);
the Foundation of the Key Laboratory of Orthopedics and
Traumatology of Traditional Chinese Medicine and Rehabilitation, Ministry of Education,
Fujian University of Traditional Chinese Medicine (grant no. XGS2024004);
and the K.~C. Wong Magna Fund at Ningbo University.
\section*{CRediT authorship contribution statement}
\textbf{Ye Ma:} Conceptualization, Methodology, Formal analysis,
Writing -- original draft, Writing -- review \& editing.
\textbf{Dongwei Liu:} Formal analysis, Writing -- original draft,
Writing -- review \& editing.
\textbf{Meijin Hou:} Conceptualization, Methodology, Investigation,
Formal analysis, Writing -- original draft, Writing -- review \& editing.
\textbf{Chenyi Guo:} Writing -- review \& editing.
\section*{Declaration of competing interest}
The authors declare that they have no competing interests.
\section*{Data availability}
The lower-limb biomechanics dataset described by \citet{camargo2021} is
publicly available through the Georgia Tech EPIC Lab:
\url{https://www.epic.gatech.edu/opensource-biomechanics-camargo-et-al/}.
The raw Tai Chi surface-electromyography data are not publicly available
to protect participant privacy but are available from the corresponding
authors on reasonable request.
\section*{Acknowledgements}
We thank the participants and research team who contributed to the original
Tai Chi sEMG recordings, and the researchers who shared the walking dataset.

\bibliographystyle{elsarticle-harv}
\bibliography{references}

\clearpage
\input{appendix_content}

\end{document}

%% file: appendix_content.tex
\begin{center}
{\Large\bfseries Appendix}\par\medskip
{\large Reconstruction quality and restart repeatability are insufficient indicators
of muscle synergy recovery by nonnegative matrix factorisation}
\end{center}

This appendix specifies signal construction, nonnegative matrix factorisation
(NMF), outcome measures and statistical analyses for 4320 synthetic surface
electromyography (sEMG) sequences and reference signals from five movement
conditions. Each synthetic sequence was an independent observation; repeated
NMF initialisations assessed optimisation variability within a sequence.

\setcounter{table}{0}
\renewcommand{\theHtable}{appendix.\arabic{table}}
\appendix
\renewcommand{\thetable}{A\arabic{table}}
\section{Synthetic sEMG signal generation}
\label{app:generator}
\subsection{Signal model and notation}
We represented each synthetic surface electromyography (sEMG) envelope
matrix as
\begin{equation}
V=W\cdot C+n.
\label{app:eq:signal-generation}
\end{equation}
Here $V\in\mathbb R_+^{M\times T}$ contains $M$ muscle channels and $T$
sample columns; $W\in\mathbb R_+^{M\times k}$ is the generating spatial
weight matrix; and $C\in\mathbb R_+^{k\times T}$ is the generating
activation coefficient matrix. Column $w_j$ of $W$ specifies the muscle
weights of component $j$, and row $c_j$ of $C$ contains its time-varying
activation coefficients. The noiseless entry for muscle $m$ at sample $t$
is $(W\cdot C)_{mt}=\sum_{j=1}^{k}W_{mj}C_{jt}$.
The ground-truth rank $k$ counts these components; it does not denote the
numerical matrix rank of the noisy input. The disturbance matrix
$n\in\mathbb R^{M\times T}$ is the realised disturbance,
including the effect of truncation at zero.

Each sequence comprised $M=16$ muscles and $L$ trials of 100 time-normalised
samples. Concatenation along the sample dimension gave $T=100\times L$
columns. One generating spatial weight matrix was shared across all trials
in a sequence, while activation coefficients varied between trials.
Superscript 0 denotes base factors before mixing; $W$ and $C$ denote the
final generating matrices. Hats denote estimates obtained by NMF.

\subsection{Noise scaling and truncation at zero}
After constructing the final $W$ and $C$, we generated independent standard
Gaussian samples and centred them over all $M\times T$ entries:
\begin{equation}
Z_{mt}\stackrel{\mathrm{iid}}{\sim}\mathcal N(0,1),\qquad
Z_0=Z-\bar Z\mathbf1.
\end{equation}
Here $\mathcal N(0,1)$ denotes a Gaussian distribution with zero mean and
unit variance, $\bar Z$ is the grand mean of $Z$, and $\mathbf1$ is an
all-ones matrix of the same dimensions. The centred samples were scaled to
the specified signal-to-noise ratio $\eta$:
\begin{equation}
n_{\mathrm{pre}}=Z_0\sqrt{\frac{\operatorname{Var}(W\cdot C)}
{\eta\operatorname{Var}(Z_0)}}.
\label{app:eq:noise-scaling}
\end{equation}
Variances were calculated over all matrix entries. This scaling sets
$\eta=\operatorname{Var}(W\cdot C)/\operatorname{Var}(n_{\mathrm{pre}})$
before truncation. The ratio is linear, rather than expressed in decibels;
a larger value indicates less noise relative to the noiseless signal.

Negative envelope values were then truncated at zero. Consequently, the
realised disturbance in Eq.~\eqref{app:eq:signal-generation} was
\begin{equation}
n=\max(0,W\cdot C+n_{\mathrm{pre}})-W\cdot C.
\label{app:eq:realised-noise}
\end{equation}
The maximum is applied elementwise. Thus $V=\max(0,WC+n_{\mathrm{pre}})$;
the specified signal-to-noise ratio applies before truncation and may differ
from the realised ratio.

\subsection{Base spatial muscle weights}
Each column $w_j^0$ of the base spatial weight matrix $W^0$ received positive Gamma-distributed background weights,
one anchor muscle unique to that component and two randomly selected secondary
muscles. Anchor muscles were assigned by a random permutation of the 16 channels;
secondary muscles were sampled without replacement from the remaining channels.
Background weights followed $\operatorname{Gamma}(0.45,0.35)$, with shape and
scale given in that order. Anchor and secondary increments followed
$\mathcal U(2.5,3.5)$ and $\mathcal U(0.35,0.9)$, respectively, where
$\mathcal U(a,b)$ denotes a uniform distribution on $[a,b]$.
Each column was normalised to unit Euclidean length using
$\operatorname{unit}(w)=w/\|w\|_2$ for a nonzero vector $w$.

A candidate column was accepted if its cosine similarity with every previously
accepted column was at most 0.65. If none of 2000 candidates met this criterion,
the candidate with the smallest maximum cosine similarity to the accepted
columns was retained. If the resulting matrix had deficient numerical column
rank, 0.25 was added to each component's anchor weight and all columns were
renormalised.

The cosine criterion applied to the base columns before spatial mixing;
it was not a constraint on the final generating weights.

\subsection{Base activation coefficients}
For component $j\in\{1,\ldots,k\}$ in trial $\ell\in\{1,\ldots,L\}$, the base activation coefficient was
\[
\begin{aligned}
c^0_{j\ell}(\tau)&=A_{j\ell}\{G(\tau;\mu_{j\ell},\sigma_{j\ell})
+\lambda_{j\ell}G(\tau;\nu_{j\ell},\omega_{j\ell})+f_{j\ell}\},\\
G(\tau;\mu,\sigma)&=\exp\{-\tfrac12[(\tau-\mu)/\sigma]^2\}.
\end{aligned}
\]
Here $A$ is the amplitude, $\mu$ and $\nu$ are the main- and secondary-burst
centres, $\sigma$ and $\omega$ are their widths, $\lambda$ scales the
secondary burst and $f$ is a positive floor. Trial phases were
$\tau=(h-0.5)/100$ for within-trial sample index $h=1,\ldots,100$.

Nominal main-burst centres were evenly spaced from 0.08 to 0.92 across
components. Independent zero-mean Gaussian timing perturbations comprised a
shared trial effect with standard deviation 0.012 and a component-specific
effect with standard deviation 0.018. Main-burst centres were clipped to
$[0.03,0.97]$. Secondary-burst centres were offset in either direction, with
equal probability, by $\mathcal U(0.18,0.34)$ and clipped to $[0.02,0.98]$.
The remaining parameters followed Table~\ref{tab:activation-parameters}.
\begin{table}[!tbp]\centering\small
\begin{tabular}{@{}ll@{}}\toprule
Parameter & Distribution\\\midrule
Log amplitude, $\log A$ & $\mathcal N(0,0.25^2)$\\
Main-burst width, $\sigma$ & $\mathcal U(0.045,0.085)$\\
Secondary-burst width, $\omega$ & $\mathcal U(0.06,0.12)$\\
Secondary-burst scale, $\lambda$ & $\mathcal U(0.08,0.22)$\\
Positive floor, $f$ & $\mathcal U(0.003,0.015)$\\\bottomrule
\end{tabular}
\caption{Distributions of base activation parameters. $\mathcal N$ denotes a
Gaussian distribution; $\mathcal U$ denotes a uniform distribution.}
\label{tab:activation-parameters}
\end{table}
Trials were concatenated to form $C^0$, with global sample index
$t=100(\ell-1)+h$.

\subsection{Factorial design and mixing procedures}
The factorial design crossed ground-truth rank $k\in\{3,5,7,9\}$;
signal-to-noise ratio $\eta\in\{0.5,2,8,32\}$; sequence length
$T\in\{300,1500,8000\}$; spatial mixing coefficient $b_W\in\{0,0.35,0.65\}$;
and activation mixing coefficient $b_C\in\{0,0.55,0.85\}$.
The full factorial design gave $4\times4\times3\times3\times3=432$
conditions. These sequence lengths corresponded to $L=3,15,80$ trials, respectively.
Ten independently seeded sequences per condition yielded
$N=4320$ sequences. Each sequence was the statistical unit; its trials
shared generating spatial weights and were not separate statistical units.

Spatial mixing moved each base muscle-weight vector towards the normalised
mean of the base vectors:
\begin{equation}
\bar w^0=\operatorname{unit}\left(k^{-1}\sum_{j=1}^{k} w_j^0\right),\qquad
w_j=\operatorname{unit}\{(1-b_W)w_j^0+b_W\bar w^0\}.
\label{app:eq:blend-w}
\end{equation}
Larger $b_W$ increased similarity among ground-truth muscle-weight vectors
while preserving unit column norms.

Activation mixing operated on the rows of $C^0$ across all concatenated
trials. Each row was moved towards the across-component mean, then rescaled
to restore its original root mean square (RMS):
\begin{equation}
\begin{aligned}
\bar c^0&=k^{-1}\sum_{j=1}^{k} c_j^0,\qquad
\widetilde c_j=(1-b_C)c_j^0+b_C\bar c^0,\\
c_j&=\widetilde c_j\,
\frac{\operatorname{RMS}(c_j^0)}{\operatorname{RMS}(\widetilde c_j)},\qquad
\operatorname{RMS}(c)=\sqrt{T^{-1}\sum_{t=1}^{T}c_t^2}.
\end{aligned}
\label{app:eq:blend-c}
\end{equation}
Larger $b_C$ increased activation overlap while preserving each row's RMS.
Zero mixing coefficients retained the base matrices. The final matrices were
checked for nonnegativity, positive component energy and numerical rank $k$
before noise addition.

Sequence seeds were derived from SHA-256 hashes of the base seed 20260905,
the identifier \texttt{unified}, the five factor levels and replicate index
0--9, joined by colons. The first four digest bytes, interpreted as an unsigned
little-endian integer, seeded NumPy's default random-number generator. Rank,
sequence length and replicate were represented as integers; the other factor
levels as floating-point values.

\section{NMF estimation, rank selection and consensus estimation}
\label{app:estimation}
\subsection{NMF estimation and rank selection}
NMF minimised $\tfrac12\|V-\widehat W_r\cdot\widehat C_r\|_F^2$ subject to
$\widehat W_r\in\mathbb R_+^{M\times r}$ and
$\widehat C_r\in\mathbb R_+^{r\times T}$, where $\|\cdot\|_F$ denotes the
Frobenius norm. Candidate rank $r$ denotes the number of fitted components.
At every rank $r\in\{1,\ldots,16\}$, coordinate descent in
scikit-learn 1.9.0 (Python 3.12.13; NumPy 2.5.2; SciPy 1.18.1) used 20
initialisations: nonnegative double singular value decomposition (NNDSVD)
with seed 0 and random initialisations with seeds 1--19. Each fit used
tolerance $10^{-4}$ and at most 500 iterations.

The fit with the smallest finite raw sum of squared errors (SSE), calculated
in double precision, supplied the paired factors $\widehat W_r$ and
$\widehat C_r$. Ties were resolved in the listed order of initialisations.
Generating factors were not used to select this fit. Its reconstructed sEMG
envelope matrix $\widehat V_r=\widehat W_r\cdot\widehat C_r$ was used to
calculate globally centred variance accounted for (VAF):
\[
v_r=1-\frac{\|V-\widehat V_r\|_F^2}{\|V-\bar V\mathbf1\|_F^2},
\]
VAF was calculated separately for each sequence; $\bar V$ is the mean
across all its muscles and time points, including concatenated trials, and
$\mathbf1$ is an all-ones matrix of the same dimensions. Reconstruction quality and spatial comparisons at a
rank used the same minimum-SSE fit. The selected rank $\widehat k$ was the
smallest $r$ with $v_r\geq0.90$, or 16 if none reached the threshold.
All candidate ranks and 20 initialisations were evaluated. Finite fits were
retained at the 500-iteration limit even if a convergence warning occurred.

\subsection{Consensus estimation for restart repeatability}
At each distinct selected or ground-truth rank, the 20 fitted spatial weight
matrices were used to calculate a separate diagnostic of restart repeatability.
Their unit-length columns were pooled and clustered by $k$-means with $r$
clusters, ten initialisations and seed 0. Normalised cluster centroids formed
the consensus spatial weight matrix $W_r^{\mathrm{cons}}$. This consensus
was used only to quantify restart repeatability, not to select a fit or to
score spatial similarity against the generating factors. Coincident selected
and ground-truth ranks reused the same fitted matrices and consensus.

\section{Assessment of rank-selection accuracy, spatial similarity and synergy recovery}
\label{app:assessment}
\subsection{Muscle-weight vector matching and outcome measures}
Spatial similarity described similarity between estimated and generating
muscle-weight vectors after one-to-one matching, quantified by spatial cosine
similarity; it did not require correct rank selection.
For ground-truth muscle-weight vector $w_j$ and minimum-SSE estimated
muscle-weight vector $\widehat w_{r,a}$, spatial cosine similarity was
\[
s_{ja}=\operatorname{unit}(w_j)^{\mathsf T}
\operatorname{unit}(\widehat w_{r,a}).
\]
A column with Euclidean norm at or below $10^{-12}$ was treated as zero,
and every cosine involving that column was defined as zero. Such a finite
degenerate fit remained eligible for minimum-SSE selection and in the
synergy recovery denominator. Optimal one-to-one assignment maximised the total spatial
cosine similarity across one-to-one matches, giving a set $\mathcal P_r$
of $m_r=\min(k,r)$ matched pairs. Unmatched components were excluded from
spatial similarity summaries. Mean spatial cosine similarity and minimum
matched spatial cosine similarity were, respectively,
\[
s_W(r)=\frac{1}{m_r}\sum_{(j,a)\in\mathcal P_r}s_{ja},\qquad
s_W^{\min}(r)=\min_{(j,a)\in\mathcal P_r}s_{ja}.
\]
Rank error was $e=\widehat k-k$: negative, zero and positive values denoted
rank under-selection, exact rank selection and rank over-selection,
respectively. Rank-selection accuracy was the percentage of sequences for
which the selected rank equalled the ground-truth rank. Synergy recovery was the
percentage of evaluated sequences for which the evaluated rank equalled the
ground-truth rank and the minimum matched spatial cosine similarity was at least
0.80. The per-sequence indicator was
\[
R_{kW}(r)=\mathbf1\{r=k\}\,\mathbf1\{s_W^{\min}(r)\geq0.80\},
\]
where $\mathbf1\{\cdot\}$ is an indicator equal to one when the stated
condition holds and zero otherwise. Selected-rank assessment used
$r=\widehat k$, whereas supplied-rank assessment used $r=k$. The rank requirement prevented unmatched components from satisfying the
recovery criterion. Recovery assessed component number and spatial weights,
with one outcome per sequence and rank setting.

Restart repeatability $A_W(r)$ quantified agreement of each of the 20 fits
with the consensus. If $a_{jb}$ is the spatial cosine similarity for
consensus component $j$ matched to fit $b$ by optimal one-to-one assignment, then
\[
A_W(r)=\frac{1}{20r}\sum_{b=0}^{19}\sum_{j=1}^{r}a_{jb}.
\]
This internal agreement measure used a fixed input and rank and included each
scored initialisation in the consensus. Missing initialisations or zero spatial
columns invalidated repeatability estimates, which were excluded from
repeatability summaries and associations without excluding the corresponding
sequences from recovery assessment.
No selected- or supplied-rank repeatability estimate was invalid.

\subsection{Statistical assessment}

\paragraph{Association analyses}\label{app:association-details}
Spearman's $\rho$ descriptively summarised the associations of reconstruction
quality and valid restart repeatability with mean spatial cosine similarity.
Secondary pooled ordinary least squares regressions used
heteroscedasticity-consistent (HC3) covariance estimates for slope tests and
confidence intervals of the fitted mean. Their two slope tests formed one
exploratory Benjamini--Hochberg (BH) false discovery rate adjustment family.
Condition-centred descriptive regressions subtracted the corresponding
condition mean from each predictor and response, distinguishing variation
within conditions from pooled variation across the design.

Condition-centred slopes were 0.400 (95\% bootstrap confidence interval
[0.320, 0.480]) for VAF and $-0.022$ [$-0.060$, 0.013] for restart
repeatability. These supplement the pooled associations reported in the main text.

Unless otherwise specified, pointwise 95\% percentile confidence intervals
used 5000 bootstrap resamples (seed 20260925), drawing ten sequences with
replacement within each of the 432 conditions. Paired records preserved
predictor--response or selected--supplied rank pairing, and condition means
were recalculated after resampling for condition-centred slopes. Intervals
were conditional on the fixed simulation design and were not simultaneous.
Adjusted values below 0.05 defined statistical significance within each
declared test family; $p_{\mathrm{BH}}$ denotes a BH-adjusted $p$ value, and
$p_{\mathrm{Holm}}$ denotes a Holm-adjusted $p$ value.

\section{Factor effects on synergy recovery}
\label{app:factor-statistics}
The outcome was the binary selected-rank synergy recovery indicator,
$R_{kW}(\widehat k)$. Main effects compared equally weighted marginal
synergy recovery probabilities, averaging over the other four factors; two-factor interactions
averaged over the other three. In condition $u$, $\bar Y_u$ was the mean of
ten independent binary synergy recovery outcomes and $s_u^2$ was their sample
variance. With $\bar Y$ collecting the 432 condition means, the estimated
covariance was $\widehat\Sigma=\operatorname{diag}(s_u^2/10)$.
The cell-mean Wald statistic was
\[
Q=(\mathcal L\bar Y)^{\mathsf T}
(\mathcal L\widehat\Sigma\mathcal L^{\mathsf T})^+
(\mathcal L\bar Y).
\]
The contrast matrix $\mathcal L$ combined equal-condition averaging with
Helmert contrasts for main effects or Kronecker products of Helmert contrasts
for interactions. Superscript $+$ denotes the Moore--Penrose pseudoinverse.
Eigenvalues at or below $10^{-10}\max(\lambda_{\max},10^{-30})$ were discarded,
where $\lambda_{\max}$ is the largest eigenvalue of the contrast covariance.
The asymptotic reference distribution was chi-squared with degrees of freedom
equal to the retained covariance rank. Five interaction covariances had
reduced rank (Table~\ref{tab:factor-diagnostics}). The norm of the observed
contrast projected onto discarded covariance directions was checked to be
at most $10^{-8}$ in every test.

The five main effects and ten interactions formed one 15-test family for
BH adjustment of the Wald $p$ values. After a significant main effect,
pairwise level comparisons used the same cell-mean covariance and Holm
adjustment within that factor. The observed difference between marginal
probabilities quantified a pairwise effect. For omnibus effects, magnitude
was summarised by the root mean square (RMS) deviation of marginal
probabilities from their overall mean for main effects, or by the RMS
interaction residual after removal of row and column means for interactions.
Multiplication by 100 expressed these effects in percentage points (pp).

Pointwise 95\% percentile confidence intervals for marginal probabilities
and pairwise differences used 9999 within-condition bootstrap resamples
(seed 20260924). Resampled synergy recovery counts followed
$\operatorname{Binomial}(10,\bar Y_u)$, equivalent to resampling the ten
observed binary outcomes within each condition. The same resampled condition
means were used for each marginal contrast, preserving the design. Conditions
with all-zero or all-one observed outcomes retain zero empirical variability
under resampling; zero-width intervals do not establish certainty about an
underlying synergy recovery probability. Finite-sample Type I error rates and
confidence-interval coverage were not evaluated.

Tables~\ref{tab:factor-diagnostics}--\ref{tab:factor-pairwise} report omnibus
tests, marginal recovery probabilities and pairwise comparisons. Marginal
effects should be interpreted alongside the interactions.

The recovery criterion, VAF-based rank-selection rule and candidate rank range
were revised after initial analyses and were not preregistered. Analyses were
therefore exploratory; ten sequences per condition and sparse recovery outcomes
limit interpretation of effects and uncertainty.

\begin{table}[!htbp]\centering\small
\setlength{\tabcolsep}{4pt}\renewcommand{\arraystretch}{1.1}
\begin{tabular}{@{}lrrrr@{}}\toprule
Effect & RMS (pp) & $Q$ & df (nom.) & $p_{\mathrm{BH}}$\\\midrule
$k$ & 7.82 & 2781.822 & 3 (3) & $<0.001$\\
$\eta$ & 8.99 & 2813.151 & 3 (3) & $<0.001$\\
$T$ & 0.69 & 18.860 & 2 (2) & $<0.001$\\
$b_W$ & 5.08 & 1244.663 & 2 (2) & $<0.001$\\
$b_C$ & 6.96 & 3030.007 & 2 (2) & $<0.001$\\
$k$ $\times$ $\eta$ & 8.59 & 3441.494 & 6 (9) & $<0.001$\\
$k$ $\times$ $T$ & 0.54 & 26.405 & 6 (6) & $<0.001$\\
$k$ $\times$ $b_W$ & 4.52 & 1277.568 & 6 (6) & $<0.001$\\
$k$ $\times$ $b_C$ & 6.45 & 3660.174 & 5 (6) & $<0.001$\\
$\eta$ $\times$ $T$ & 0.85 & 23.446 & 5 (6) & $<0.001$\\
$\eta$ $\times$ $b_W$ & 5.26 & 1514.715 & 5 (6) & $<0.001$\\
$\eta$ $\times$ $b_C$ & 7.56 & 3271.147 & 5 (6) & $<0.001$\\
$T$ $\times$ $b_W$ & 0.52 & 9.472 & 4 (4) & 0.0503\\
$T$ $\times$ $b_C$ & 0.60 & 22.485 & 4 (4) & $<0.001$\\
$b_W$ $\times$ $b_C$ & 3.96 & 1262.015 & 4 (4) & $<0.001$\\
\bottomrule\end{tabular}
\caption{Cell-mean Wald tests of five main effects and ten interactions on selected-rank synergy recovery.}
\label{tab:factor-diagnostics}

\par\smallskip\raggedright\footnotesize
$k$: ground-truth rank; $\eta$: signal-to-noise ratio; $T$: sequence length;
$b_W$: spatial mixing coefficient; $b_C$: activation mixing coefficient.
RMS: root mean square effect; pp: percentage points; df: retained covariance
rank; nom.: nominal contrast dimension. BH: Benjamini--Hochberg adjustment
within the 15-test family. Each test uses 432 conditions with ten independent
sequences per condition.
\end{table}

\begin{table}[!htbp]\centering\small
\setlength{\tabcolsep}{5pt}
\begin{tabular}{@{}llrrl@{}}\toprule
Factor & Level & $n$ & \shortstack{Recovery\\count} & \shortstack{Synergy recovery (\%)\\{[95\% CI]}}\\\midrule
$k$ & 3 & 1080 & 210 & 19.4 [18.6, 20.2]\\
 & 5 & 1080 & 116 & 10.7 [10.1, 11.4]\\
 & 7 & 1080 & 15 & 1.4 [0.9, 1.9]\\
 & 9 & 1080 & 1 & 0.1 [0.0, 0.3]\\
\addlinespace
$\eta$ & 0.5 & 1080 & 0 & 0.0 [0.0, 0.0]\\
 & 2 & 1080 & 1 & 0.1 [0.0, 0.3]\\
 & 8 & 1080 & 237 & 21.9 [21.0, 22.9]\\
 & 32 & 1080 & 104 & 9.6 [9.0, 10.3]\\
\addlinespace
$T$ & 300 & 1440 & 100 & 6.9 [6.4, 7.5]\\
 & 1500 & 1440 & 120 & 8.3 [7.8, 8.8]\\
 & 8000 & 1440 & 122 & 8.5 [8.0, 8.9]\\
\addlinespace
$b_W$ & 0 & 1440 & 201 & 14.0 [13.3, 14.6]\\
 & 0.35 & 1440 & 119 & 8.3 [7.8, 8.8]\\
 & 0.65 & 1440 & 22 & 1.5 [1.2, 1.9]\\
\addlinespace
$b_C$ & 0 & 1440 & 244 & 16.9 [16.3, 17.6]\\
 & 0.55 & 1440 & 98 & 6.8 [6.2, 7.4]\\
 & 0.85 & 1440 & 0 & 0.0 [0.0, 0.0]\\
\bottomrule\end{tabular}
\caption{Equally weighted marginal selected-rank synergy recovery by factor level.}
\label{tab:factor-marginals}

\par\smallskip\raggedright\footnotesize
$n$: number of sequences; CI: pointwise percentile confidence interval from 9999 within-condition
bootstrap resamples. Factor symbols are defined in
Table~\ref{tab:factor-diagnostics}. Recovery count: sequences meeting the exact-rank
and minimum matched spatial cosine similarity criterion. Marginals average
equally over all levels of the other four factors.
\end{table}

\begin{table}[!htbp]\centering\small
\setlength{\tabcolsep}{4pt}
\begin{tabular}{@{}llrr@{}}\toprule
Factor & Levels & \shortstack{Difference (pp)\\{[95\% CI]}} & $p_{\mathrm{Holm}}$\\\midrule
$k$ & 5 minus 3 & -8.70 [-9.72, -7.69] & $<0.001$\\
 & 7 minus 3 & -18.06 [-18.98, -17.13] & $<0.001$\\
 & 9 minus 3 & -19.35 [-20.09, -18.52] & $<0.001$\\
 & 7 minus 5 & -9.35 [-10.19, -8.52] & $<0.001$\\
 & 9 minus 5 & -10.65 [-11.30, -10.00] & $<0.001$\\
 & 9 minus 7 & -1.30 [-1.85, -0.83] & $<0.001$\\
\addlinespace
$\eta$ & 2 minus 0.5 & 0.09 [0.00, 0.28] & 0.317\\
 & 8 minus 0.5 & 21.94 [21.02, 22.87] & $<0.001$\\
 & 32 minus 0.5 & 9.63 [8.98, 10.28] & $<0.001$\\
 & 8 minus 2 & 21.85 [20.93, 22.78] & $<0.001$\\
 & 32 minus 2 & 9.54 [8.89, 10.19] & $<0.001$\\
 & 32 minus 8 & -12.31 [-13.43, -11.20] & $<0.001$\\
\addlinespace
$T$ & 1500 minus 300 & 1.39 [0.69, 2.08] & $<0.001$\\
 & 8000 minus 300 & 1.53 [0.83, 2.22] & $<0.001$\\
 & 8000 minus 1500 & 0.14 [-0.49, 0.76] & 0.691\\
\addlinespace
$b_W$ & 0.35 minus 0 & -5.69 [-6.46, -4.93] & $<0.001$\\
 & 0.65 minus 0 & -12.43 [-13.12, -11.74] & $<0.001$\\
 & 0.65 minus 0.35 & -6.74 [-7.36, -6.18] & $<0.001$\\
\addlinespace
$b_C$ & 0.55 minus 0 & -10.14 [-11.04, -9.31] & $<0.001$\\
 & 0.85 minus 0 & -16.94 [-17.57, -16.32] & $<0.001$\\
 & 0.85 minus 0.55 & -6.81 [-7.36, -6.25] & $<0.001$\\
\bottomrule\end{tabular}
\caption{Within-factor differences in marginal selected-rank synergy recovery.}
\label{tab:factor-pairwise}

\par\smallskip\raggedright\footnotesize
Contrasts subtract the lower listed level from the higher listed level.
pp: percentage points; CI: pointwise percentile confidence interval from 9999
within-condition bootstrap resamples. Holm adjustment was applied separately
within each factor. Each ground-truth-rank or signal-to-noise-ratio level uses
1080 sequences; each level of the remaining factors uses 1440 sequences.
Factor symbols are defined in Table~\ref{tab:factor-diagnostics}.
\end{table}

\clearpage
\section{Synergy recovery with the ground-truth rank supplied}
\label{app:supplied-rank}
Each sequence was assessed at its selected and ground-truth ranks using the
same minimum-SSE fitting and matching procedures. At the supplied rank,
recovery required minimum matched spatial cosine similarity $\geq0.80$.
The paired difference in recovery indicators was averaged over all sequences;
95\% confidence intervals used 5000 within-condition bootstrap resamples
(seed 20260925), preserving the pairing between rank settings.

Recovery increased from 342/4320 sequences (7.9\%, 95\% confidence interval
[7.6, 8.2]) to 2626/4320 (60.8\% [60.0, 61.5]), a paired gain of
52.9 percentage points [52.1, 53.6]. Corresponding paired changes were
$-0.0631$ [$-0.0634$, $-0.0628$] for VAF and 0.0259 [0.0250, 0.0267]
for mean spatial cosine similarity.

\section{Evaluation with measured sEMG signals}
\label{app:reference-signals}
\subsection{Reference-factor construction}

Walking recordings came from the dataset of Camargo et al. (2021;
\href{https://doi.org/10.1016/j.jbiomech.2021.110320}{doi:10.1016/j.jbiomech.2021.110320}).
The reference-construction procedure applied zero-phase fourth-order
20--450 Hz band-pass filtering, a 60 Hz notch (quality factor 30), full-wave
rectification and fourth-order 6 Hz low-pass filtering. Envelopes were
interpolated to 100 samples per recording using natural cubic splines, with
negative values truncated at zero, and normalised by each participant's
muscle-specific peak. Recordings failing checks for finite values, flat channels,
severe clipping or mains interference were excluded; equal numbers of retained
recordings per participant were used within each walking condition.
Tai Chi reference construction used existing nonnegative envelopes,
time-normalised to 125 samples and normalised by participant-specific muscle
peaks. The reference signals contained 270, 540 and 315 walking recordings and
174 and 190 Heel Kick recordings for left and right stance, respectively.

Reference ranks were established during reference construction, before the
present recovery assessment. The construction procedure selected the smallest
rank reaching VAF 0.90; otherwise it used the first rank for which adding a
component increased VAF by less than 0.05, or the maximum candidate rank if
neither criterion was met. Construction considered ranks up to nine for
walking and ten for Tai Chi, using at most 8000 sampled columns for rank
selection. Tai Chi spatial weights were shared consensus estimates from
recordings pooled within each stance condition. These reference-construction
settings are distinct from the rank-selection procedure subsequently evaluated
on the noiseless reference signals.

Five fixed reference signals were constructed from spatial weight matrices
$W_{\mathrm{ref}}$ estimated from measured sEMG. The reference rank
$k_{\mathrm{ref}}$ is the number of components in each reference signal.
Level, ramp and stair walking used ten muscle channels and reference ranks
$k_{\mathrm{ref}}=6,6,7$, respectively. Tai Chi Heel Kick with left and right
stance used 16 channels and reference ranks of six each. Walking reference
weights were consensus estimates from 20 fits; Tai Chi reference weights
were shared across recordings of the same movement and stance side. Holding
$W_{\mathrm{ref}}$ fixed, nonnegative least squares estimated activation
coefficients from each measured envelope block. Concatenating these blocks
gave $C_{\mathrm{ref}}$ and the noiseless reference signal
$E_{\mathrm{syn}}=W_{\mathrm{ref}}\cdot C_{\mathrm{ref}}$.
The signals contained 27000, 54000, 31500, 21750 and 23750 time points for
level walking, ramp walking, stair walking, left stance and right stance,
respectively. These spatial weights and activation coefficients defined the known reference
structure for the subsequent recovery assessment.

\subsection{Rank selection, synergy recovery and restart repeatability}
Reference signals were evaluated using the NMF, VAF and matching procedures
in \ref{app:estimation} and~\ref{app:assessment}, with candidate
ranks 1--10 for walking and 1--16 for Tai Chi. Each signal contributed one
recovery indicator at the selected rank and one at its supplied reference rank;
recovery required $r=k_{\mathrm{ref}}$ and minimum matched spatial cosine
similarity $\geq0.80$.

All five selected ranks were 5, below their reference ranks, so none met the
recovery criterion despite VAF values of 0.9182--0.9514. All five met the
criterion at their supplied reference ranks (Table~\ref{tab:reference-bestfit}).
These are descriptive results for five fixed reference signals; NMF
initialisations were not independent biological observations.

\begin{table}[!htbp]\centering\small
\setlength{\tabcolsep}{3pt}\renewcommand{\arraystretch}{1.1}
\begin{tabular}{@{}p{.21\textwidth}lrrrrrc@{}}\toprule
Signal & Context & $r$ & VAF & $s_W$ & $s_W^{\min}$ & $A_W$ & \shortstack{Recovery\\criterion}\\\midrule
Level walking & Selected & 5 & 0.9514 & 0.9781 & 0.9078 & 0.9999 & No\\
 & Supplied & 6 & 1.0000 & 0.9998 & 0.9994 & 0.9998 & Yes\\
\addlinespace
Ramp walking & Selected & 5 & 0.9327 & 0.9694 & 0.8903 & 0.9408 & No\\
 & Supplied & 6 & 1.0000 & 0.9998 & 0.9997 & 0.9999 & Yes\\
\addlinespace
Stair walking & Selected & 5 & 0.9182 & 0.9768 & 0.9427 & 1.0000 & No\\
 & Supplied & 7 & 1.0000 & 0.9980 & 0.9887 & 0.9989 & Yes\\
\addlinespace
Heel Kick, left stance & Selected & 5 & 0.9329 & 0.9906 & 0.9781 & 0.9552 & No\\
 & Supplied & 6 & 1.0000 & 1.0000 & 1.0000 & 1.0000 & Yes\\
\addlinespace
Heel Kick, right stance & Selected & 5 & 0.9397 & 0.9812 & 0.9300 & 1.0000 & No\\
 & Supplied & 6 & 1.0000 & 0.9999 & 0.9998 & 0.9998 & Yes\\
\bottomrule\end{tabular}
\caption{Recovery assessment and restart repeatability for reference signals from five movement conditions.}
\label{tab:reference-bestfit}

\par\smallskip\raggedright\footnotesize
$r$: evaluated rank; VAF: globally centred variance accounted for;
$s_W$: mean spatial cosine similarity; $s_W^{\min}$: minimum matched
spatial cosine similarity. Both spatial summaries compare matched estimated
and reference muscle-weight vectors. $A_W$: consensus restart repeatability.
Supplied: the reference rank was provided to NMF. Recovery criterion: whether
the single minimum-SSE fit met the exact-reference-rank and minimum matched
spatial cosine similarity criterion.
Spatial scores and VAF use that fit; $A_W$ uses all 20 initialisations and
their consensus. Values are rounded; all supplied-rank VAFs exceeded 0.999999.
\end{table}

All reported selected- and supplied-reference-rank fits were finite, had no
zero components and emitted no convergence warnings; all corresponding
repeatability estimates were valid. No reference-signal outcome was excluded.